\documentclass[pre, reprint, showpacs, showkeys]{revtex4-1}
\usepackage{hyperref}
\usepackage{graphicx}
\usepackage{amsmath}
\usepackage{amsfonts}

\begin{document}
\title{Rejection-free Monte-Carlo sampling for general potentials}
\author{E.A.J.F.\ Peters}
\email{e.a.j.f.peters@tue.nl}
\author{G.\ de With}
\email{g.deWith@tue.nl}
\date{\today}

\begin{abstract}
A Monte Carlo method to sample the classical configurational canonical ensemble is introduced.
In contrast to the Metropolis algorithm, where trial moves can be rejected, in this approach collisions take place.
The implementation is event-driven, i.e., at scheduled times the collisions occur.
A unique feature of the new method is that smooth potentials (instead of only step-wise changing ones) can be used.
Besides an event-driven approach where all particles move simultaneously, we also introduce a straight event-chain implementation.
As proof-of-principle a system of Lennard-Jones particles is simulated.
\end{abstract}


\maketitle

\section{Introduction}

The most commonly used methods for simulating particle systems in accordance to classical statistical mechanics are molecular dynamics (MD) and Monte-Carlo methods (MC) based on the Metropolis scheme \cite{Binder87, Allen89, Frenkel02, Rapaport04, Krauth06}.
For systems, such as hard-sphere systems, with impulsive interactions a time-driven MD approach does not work and an event-driven approach can be used.
In fact, the pioneering work of Alder and Wainwright used an event-driven molecular dynamics (ED-MD) scheme \cite{Alder59}.

In the Metropolis MC scheme trial moves are either accepted or rejected.
In highly concentrated systems the acceptance rate can be very low and
simulating using MD requires very small time-steps.
In dilute systems the time-scale in MD or step-size in MC is determined by the molecular collision process and simulation time is wastefully spend on flying through empty space.
In both cases an event-driven approach can speed up the computation.

ED-MD can be generalized to hard-spheres to potentials build up by a sequence of steps \cite{Chapela84}.
Clearly in this case an event takes place at each step.
The method we derive in this paper differs in several aspects from ED-MD: Collision-events are determined by means of a stochastic process. Potentials are not necessarily step-wise. There is no exchange of kinetic and potential energy. In fact momentum is not relevant and the configurational canonical is sampled directly.

Instead of rejecting moves as in the Metropolis scheme a collision takes place.
There is quite some freedom to model a collision event.
One possibility is to model it as a Newtonian collision.
Another possibility is to move one particle at a time where, at collision, another particle takes over.
This is similar to the straight event-chain collision in hard-sphere simulation \cite{Bernard09, Bernard11}.

On an algorithmic level there is some similarity with kinetic (or dynamic) MC \cite{Fichthorn91} and $n$-fold way MC simulations \cite{Bortz75, Guerra09}.
In these methods there is a finite number of (classes of) moves modeled as Poisson processes.
Using the rates corresponding the these Poisson processes the moment in time a next event occurs can be computed.
The efficiency of these methods is determined by the fact that the number of moves is finite which is not the case in a particle system.
Also in these kinetic MC simulations nothing happens in between two subsequent events.
In the method we will outline below, however, particles will move linearly in between subsequent collision events.
Therefore even when no events occur the system is evolving.
The present method, which is surprisingly simple, is a unique event-driven Monte Carlo method.

\section{An event driven stochastic scheme}

The prototypical Monte-Carlo scheme for sampling a configurational canonical distribution generates ``moves'' from an old state, $\mathbf{x}^n_\mathrm{old}$, to a new state, $\mathbf{x}^n_\mathrm{new}$, according to a conditional probability density $T(\mathbf{x}^n_\mathrm{new}|\mathbf{x}^n_\mathrm{old})$.
The transitional probabilities are forced to obey the detailed-balance relation,
\begin{equation}
T(\mathbf{y}^n | \mathbf{x}^n) \, \exp[ - \beta \, U(\mathbf{x}^n)] = T(\mathbf{x}^n | \mathbf{y}^n) \, \exp[ - \beta \, U(\mathbf{y}^n)].
\label{eq:transitions}
\end{equation}
In the Metropolis scheme we decompose the transition probability density as,
\begin{equation}
T(\mathbf{y}^n | \mathbf{x}^n) = \mathrm{acc}(\mathbf{y}^n , \mathbf{x}^n) \, a(\mathbf{y}^n | \mathbf{x}^n),
\end{equation}
where $a(\mathbf{y}^n | \mathbf{x}^n)$ is the probability density for generating a trial move from $\mathbf{x}^n$ to $\mathbf{y}^n$ and $\mathrm{acc}(\mathbf{y}^n , \mathbf{x}^n)$ the probability that this move is accepted.
The Metropolis form for the acceptance probability equals,
\begin{equation}
\mathrm{acc}(\mathbf{x}^n_\mathrm{new} , \mathbf{x}^n_\mathrm{old}) = \min \left(1, \exp[ - \beta \, \Delta U] \right),
\label{eq:metropolis}
\end{equation}
if $a(\mathbf{y}^n | \mathbf{x}^n)=a(\mathbf{x}^n | \mathbf{y}^n)$ $\forall \, \mathbf{x}^n, \, \mathbf{y}^n$.
When a move is not accepted the positions remain unchanged: $\mathbf{x}^n_\mathrm{new} := \mathbf{x}^n_\mathrm{old}$.

\begin{figure}
\includegraphics[width=0.8\columnwidth]{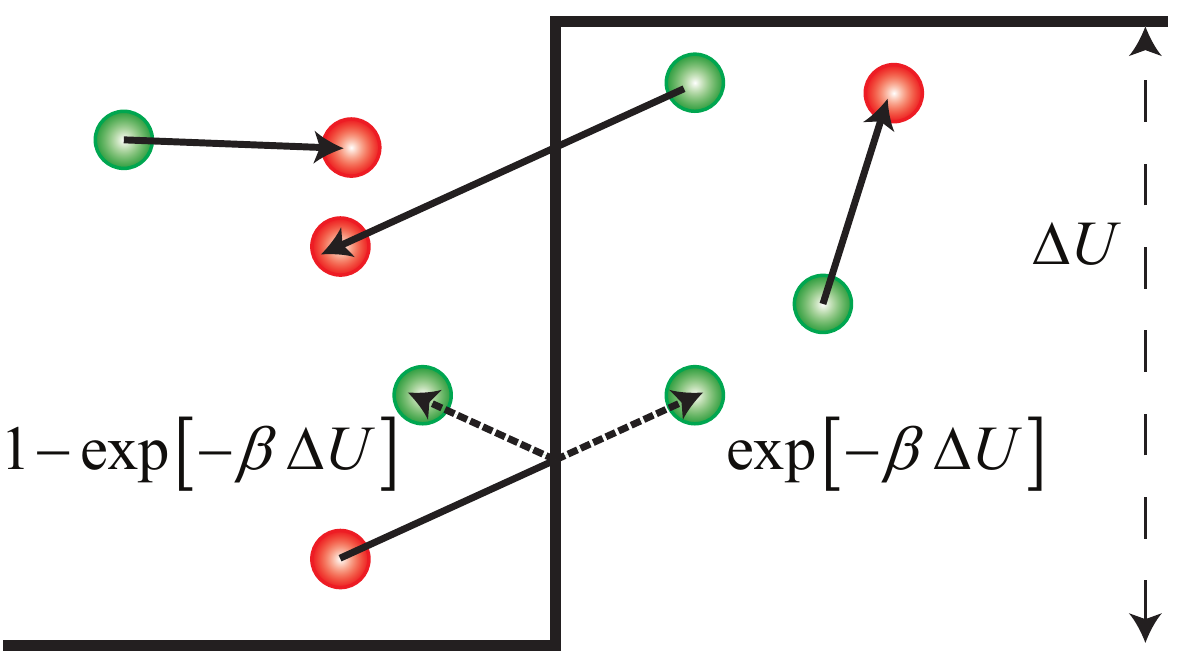}
\caption{(Color online) A trial move that moves upward to the higher energy state may give rise to a collision.}
\label{fig:collisions}
\end{figure}

Now let's consider a simple one-dimensional potential step of height $\Delta U$.
In this case detailed balance, Eq.~\eqref{eq:transitions}, can be obeyed in a different way.
Instead of rejecting a move, if a random number is below the Metropolis acceptance probability, it will collide.
So, let's consider a trial move from a position $x_\mathrm{old}$ to $x_\mathrm{new}$.
If both positions are at the same side of the barrier the move will be accepted.
If the move descents the barrier, i.e., $\Delta U < 0$ then the move is also accepted.
If the move is up the barrier, i.e., $\Delta U > 0$ it will only sometimes be accepted.
If it is not accepted, it is not rejected but the path is changed by means of a collision against the ``wall'' of the barrier (see Fig.~\ref{fig:collisions}).
Clearly a position $x_\mathrm{new}$ that is on the other side of the barrier as $x_\mathrm{old}$ can only be sampled if no collision has taken place.
For the probability that no collision occurs we use Eq.~\eqref{eq:metropolis}, $P_\mathrm{no-coll}(x_\mathrm{new} , x_\mathrm{old}) = \mathrm{acc}(x_\mathrm{new} , x_\mathrm{old})$.

Next let's consider a number of potential steps in a sequence.
For a trial move from $x_\mathrm{old}$ to $x_\mathrm{new}$ we compute the probability to not collide at each individual barrier that is crossed by means of Eq.~\eqref{eq:metropolis}.
In this case the probability to still have not experienced any collisions when reaching position $x_\mathrm{new}$ equals,
\begin{equation}
\begin{split}
  P_\mathrm{no-coll}(x_\mathrm{new} , x_\mathrm{old}) &= \prod_i \min \left(1, \exp[ - \beta \, \Delta U_i] \right)\\
  &= \exp\Bigl[ - \beta \, \sum_i \mathrm{max}(\Delta U_i,0)\Bigr],
\end{split}
\end{equation}
where the index $i$ labels the barriers crossed when moving from $x_\mathrm{old}$ to $x_\mathrm{new}$.
For every change in potential we decide to count it or not depending on the fact whether it is increasing the potential energy or not.
Going down the barrier is free, every uphill motion counts and accumulates until a collision becomes inevitable (or until the potential does not grow anymore).

We could approximate a continuous potential by a sequence of barriers and do our calculation accordingly but we will proceed differently.
If we take the limit to indefinitely small potential steps we obtain,
\begin{equation}\label{eq:P_no-coll}
 P_{\mathrm{no-coll },\alpha}(s) = \exp\left[ - \beta \, \int_{s_0}^s \mathrm{max} \Bigl (\frac{d}{d\tilde{s}} U_\alpha(\mathbf{x}^n(\tilde{s})), 0 \Bigr) \, d\tilde{s} \right],
\end{equation}
which is the conditional probability that a particle moving in a linear motion from $\mathbf{x}^n(s_0)$ to $\mathbf{x}^n(s)$ did not experience a collisions along the way.
Here we presented the formula for a general $n$-particle system, and a potential $U_\alpha$, where the subscript $\alpha$ is just a label to identify the potential (which is useful for reasons that will become apparent below).

In practice computation of the integral is trivial if one has an expression for the potential $U_\alpha(\mathbf{x}^n(s))$.
One needs to know the location maxima and minima of the potential along the path, $\mathbf{x}^n(s)$, to be able to extract increasing contributions only.
With this accumulative probability the position at which the particle does collide can be determined as follows: Draw a uniform number, $u$, between 0 and 1.
The collision takes place at the time, $s$, for which $u = P_\mathrm{no-coll}(s)$, or equivalently,
\begin{equation}\label{eq:collsion-time}
  \int_{s_0}^s \mathrm{max} \Bigl (\frac{d}{d\tilde{s}} U_\alpha\bigl(\mathbf{x}^n(\tilde{s})\bigr), 0 \Bigr) \, d\tilde{s} = -kT \ln u.
\end{equation}

\subsection{1-D proof-of-principle}

\begin{figure}
\includegraphics[width=\columnwidth]{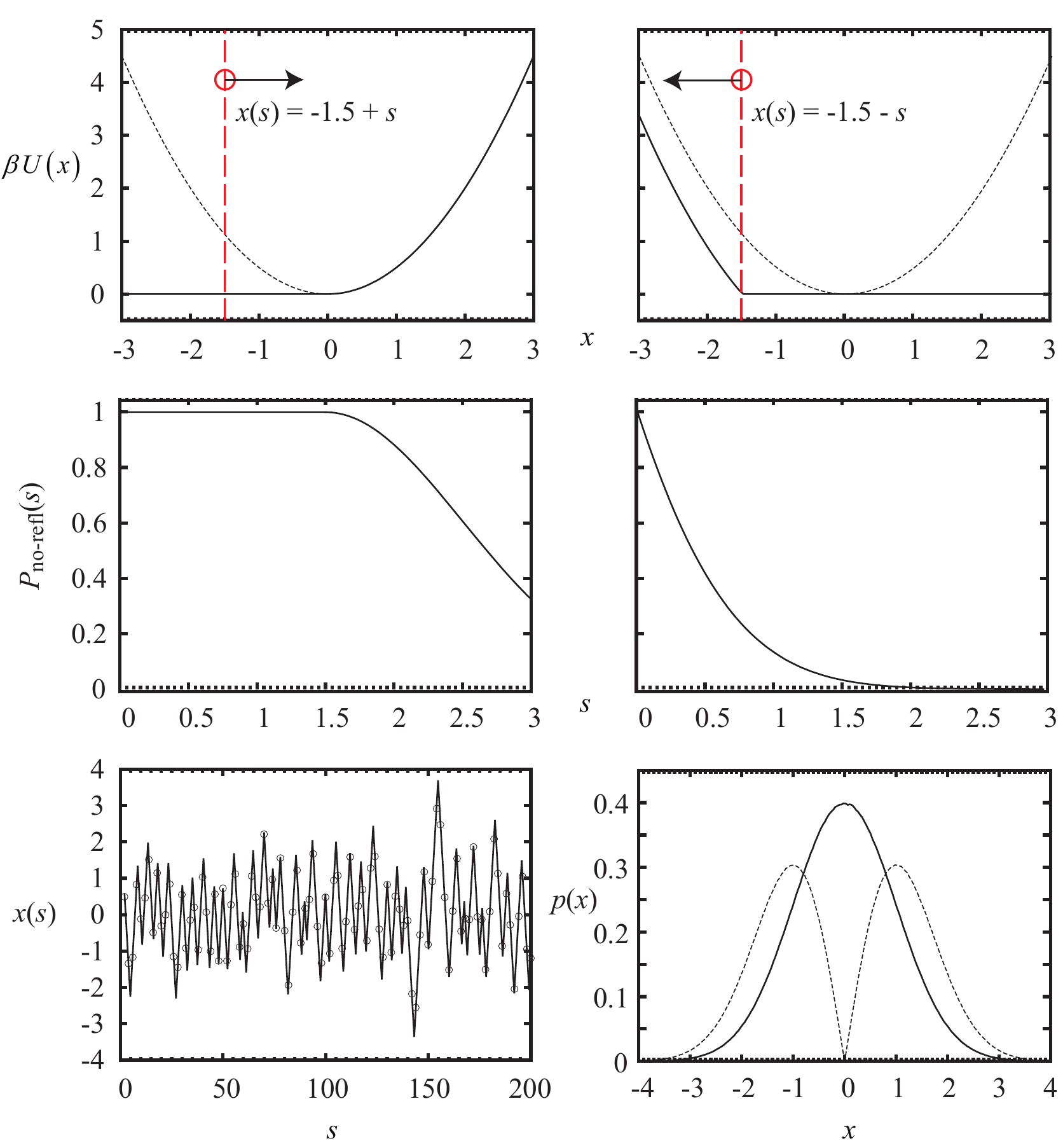}
\caption{(Color online) The upper-left graph shows the relevant part of the potential for motion to the right starting at $x=-1.5$ and the upper-right graph for motion to the left.
The cumulative probabilities to be not collide are shown in the second row of graphs.
In the lower-left corner a typical time-series is depicted.
The symbols indicate equidistantly spaced points along the $s$ axis.
These points sample the canonical ensemble as shown in the lower-right graph (solid line).
When making a histogram of the collision points one finds the dashed curve.}
\label{fig:harmonic}
\end{figure}

To prove that the scheme correctly works in practice we consider the motion in a harmonic well: $U = \tfrac{1}{2} \, x^2$.
Here we use dimensionless units $kT=1$ and the characteristic length scale equals 1.

The motion of $x$ is linear $dx/ds = v$ (constant $v$) and at a collision: $v := -v$.
Note that after the collision also $dU/ds$ has changed sign.
At this point, say at position $x_\mathrm{coll}$ and time $s_\mathrm{coll}$, we proceed with the linear motion and determine the new cumulative probability not to collide by means of Eq.~\eqref{eq:P_no-coll} and integrating from an initial position $x_\mathrm{coll}$.
Using this new cumulative probability the next collision is determined by means of solving Eq.~\eqref{eq:collsion-time} with $s_0 = s_\mathrm{coll}$.

To illustrate the process, in Fig.~\ref{fig:harmonic} the particle starts to move at $x = -1.5$. First, the collision ``time'' and position are determined, then the collision is performed by reversing the ``velocity''.
Here quotation marks are used because not ``time'', but contour length $s$ is the relevant parameter.
The ``velocity'' does not have physical significance, e.g., as used for a kinetic energy.
It is, however, more intuitive to speak in terms of time as the variable that parameterizes the path.

To generate the canonical ensemble the positions, $x$, need to be sampled at equidistant points in time.
If we define a time-step, say $\Delta s$, at every time $s_n=  n \Delta s$ the distribution is sampled.
In the lower-left graph of Fig.~\ref{fig:harmonic} the data points corresponding to $\Delta s=2$ are shown in the time-series.
When collecting these points to form a histogram the correct canonical ensemble is sampled as is shown in the lower-right graph.
It is a rigorously valid procedure, obeying detailed balance, if a new velocity $v$ is drawn from a probability distribution, which is even in $v$, at equidistantly spaced times.
In the series generated to produce the bottom graphs, however, we do not do this and just proceed along the path until the next collision occurs.
The dynamics has enough inherent randomization to cause ergodicity.
The velocities are -1 or 1 with equal statistical weight and clearly not distributed according to, e.g., a Maxwell-Boltzmann distribution.

\section{A 3-D multi-particle system}

\begin{figure}
\includegraphics[width=0.8\columnwidth]{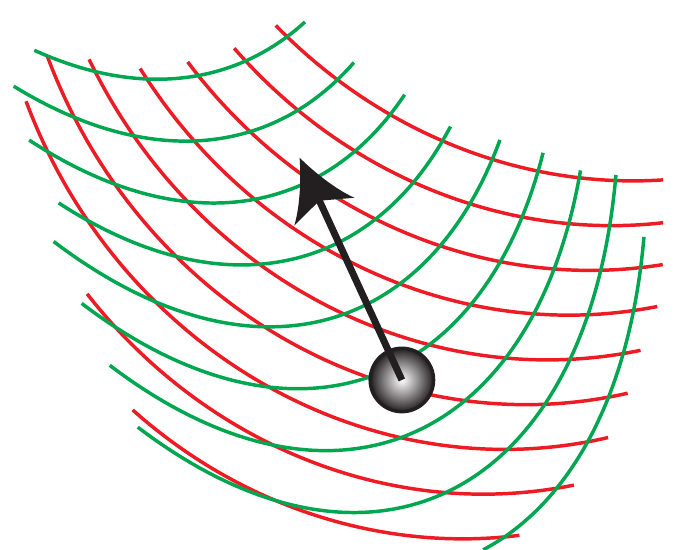}
\caption{(Color online) A particle moving in two potentials indicated by the two sets of equipotential contours.}
\label{fig:potentials}
\end{figure}

Let's consider a particle system.
Here the total potential can decomposed as a sum of potentials: $U = \sum_\alpha U_\alpha$.
In Fig.~\ref{fig:potentials} a particle moving in the fields of two potentials is shown.
Now, assume for a moment that the potentials do not increase smoothly, but stepwise at every depicted equipotential contour.
Using the same reasoning as before, at every step that is crossed by the path of the particle a collision can take place.
The probability that a path of the particle crosses a step of both potentials exactly at the same time is zero.
Therefore, in the stepwise case, it is clear that the influence of each potential $U_\alpha$ can be considered separately and
this remains valid in the limit of smooth potentials.
For each potential $U_\alpha$ individually Eq.~\eqref{eq:P_no-coll} can be used.

\subsection{Collision rules}

Let the particles in the system move with constant velocity, $\mathbf{v}^n = (\mathbf{v}_1, \mathbf{v}_2, \dots, \mathbf{v}_n)$.
If a collision due to potential $U_\alpha$ takes place at time $s_\mathrm{coll}$ the velocity after collision changes as,
\begin{equation}\label{eq:collision_rule}
    \mathbf{v}^n(s_\mathrm{coll}^+) = (\mathbf{I} - 2 \mathbf{P}_\alpha) \cdot \mathbf{v}^n(s_\mathrm{coll}^-),
\end{equation}
where $\mathbf{P}_\alpha$ is a projection matrix ($\mathbf{P}_\alpha \cdot \mathbf{P}_\alpha = \mathbf{P}_\alpha$).
A general form for the projection operator is,
\begin{equation}\label{eq:projection}
    \mathbf{P}_\alpha = \frac{\mathbf{M} \cdot \boldsymbol{\nabla} U_\alpha \, \boldsymbol{\nabla} U_\alpha}{\boldsymbol{\nabla} U_\alpha \cdot \mathbf{M} \cdot \boldsymbol{\nabla} U_\alpha}.
\end{equation}
Here the potential gradient indicates the direction normal to the equipotential surface of $U_\alpha$.
One can verify that the collision leaves the scalar $\mathbf{v}^n \cdot \mathbf{M}^{-1} \cdot \mathbf{v}^n$ invariant.

A possible simulation protocol proceeds as follows: Draw velocities from a Gaussian distribution with a co-variance matrix proportional to $\mathbf{M}$. Next run the event-driven collision scheme for a time-interval $\Delta s=1$. Lastly redraw the velocities and repeat.
This scheme gives rise to a Markov chain that obeys detailed balance.
In our simulation results we find that, in fact, the velocities do not need to be redrawn.
In appendix \ref{app:proof} we provide a proof that the algorithm indeed samples the configurational canonical ensemble.

In the case that a pair-potential acts between particles $1$ and $2$, $U_\alpha(\mathbf{x}^n) = U(\mathbf{x}_1, \mathbf{x}_2)$, the potential gradient row-vector only has non-zero entries for particles $1$ and $2$.
For the simulations we made the simple choice $\mathbf{M} = \mathbf{I}$.
For pairwise central potentials, $U_\alpha(\mathbf{x}^n) = U(|\mathbf{x}_2-\mathbf{x}_1|)$, we find $\mathbf{P}_{\alpha,\, ij} = \tfrac{1}{2} (\delta_{i1}\delta_{j1} + \delta_{i2}\delta_{j2} - \delta_{i1}\delta_{j2} - \delta_{i2}\delta_{j1}) \, \mathbf{e}_r \mathbf{e}_r$, with $\mathbf{e}_r$ the radial direction vector $\mathbf{e}_r = (\mathbf{x}_2-\mathbf{x}_1) / |\mathbf{x}_2-\mathbf{x}_1|$.
This is a formal notation equivalent to an elastic Newtonian collision between two particles of equal mass.

The simulation protocol is very similar to event-driven MD \cite{Rapaport09}.
Initially for all possible pairs a possible collision event is computed and stored in a priority queue.
If the collision that involves particles $i$ and $j$ pops up it is handled.
Now all previously computed collisions involving $i$ or $j$ become invalidated and are removed from the queue.
So, for all pairs $i-k$ and $j-k$ new collision times need to be computed similarly as in Eq.~\eqref{eq:collsion-time} by inverting Eq.~\eqref{eq:P_no-coll}.
From a computational point of view it is most efficient to perform the updating asynchronously, i.e., the particles are moved only at the moment they participate in a collision, otherwise the positions remain fixed at the spot the last collision occurred.
However, to generate the statistics we need to sample the system at equidistantly spaces time-intervals $s_\mathrm{stamp} = n \Delta s$.
We also schedule these time stamps, such that at every time $s_\mathrm{stamp}$ the positions of the particles, $\mathbf{x}^n(s_\mathrm{stamp})$, can be computed.

\subsection{Straight event-chain collisions}

It has recently been shown that straight event-chain updates can be very efficient for concentrated hard core systems \cite{Bernard09, Bernard11}.
This makes implementation simpler than for the scheme outlined above because no event-queue is needed. Hence we tested also this scheme.

If a particle, $i$, that moves with a velocity, $\mathbf{v}_i=\mathbf{v}$, collides with a particle, $j$, it stops ($\mathbf{v}_i:=0$) and the other particle takes over ($\mathbf{v}_j:= \mathbf{v}$).
The motion with collisions continue until $\Delta s=1$.
It was found that, when this scheme was performed non-reversibly, e.g., by giving particles either one out of three possible velocities: $\mathbf{v} = v \mathbf{e}_x$, $\mathbf{v} = v \mathbf{e}_y$ or $\mathbf{v} = v \mathbf{e}_z$ (and not the negative direction), the speed up was significant.
The reason is that the dynamics is non-diffusive.
Clearly in this case detailed balance is not obeyed but for hard sphere systems it was found that the correct configurational canonical ensemble is sampled.

\subsection{Lennard-Jones interaction}

\begin{figure}
\includegraphics[width=0.8\columnwidth]{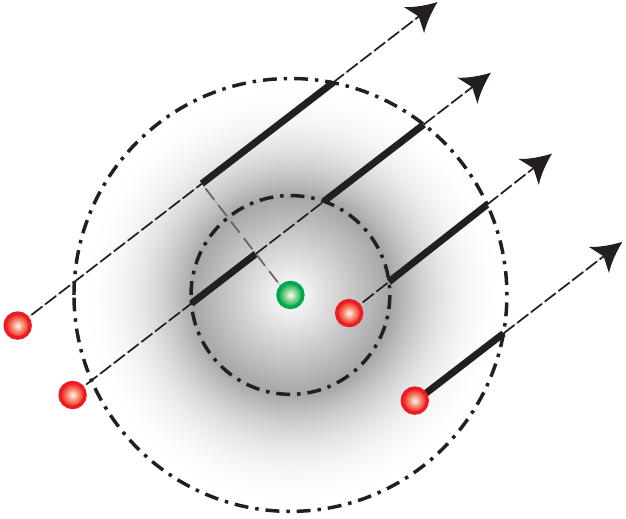}
\caption{(Color online) Particles interact with the central bead by means of a truncated-shifted Lennard-Jones interaction. The inner dash-dotted circle indicates the location of the potential minimum. The outer dot-dashed circle indicates the location of the cutoff radius. The bold solid pieces of the particle trajectories indicate the parts where the potential increases when the motion proceeds. These parts contribute in Eq.~\protect\eqref{eq:P_no-coll}. In the other sections of the paths no collision can occur.}
\label{fig:potentials_attractive}
\end{figure}

As a second example we will have a look at the truncated-shifted Lennard-Jones interaction.
\begin{equation}\label{eq:LJ}
\begin{split}
    U_\mathrm{LJ}(r) &= 4 \epsilon \left( \left( \frac{\sigma}{r}\right)^{12}-\left( \frac{\sigma}{r}\right)^6\right)\\
    U^\mathrm{trunc}_\mathrm{LJ}(r) &=
    \begin{cases}
    U_\mathrm{LJ}(r)-U_\mathrm{LJ}(r_c), & \text{for } r < r_c \\
    0, &\text{ otherwise}
    \end{cases}
\end{split}
\end{equation}

When particles move towards each other the pair-potential differences increases when it is in the repulsive regime.
Once beyond the point of closest approach, and inside the repulsive regime, the motion is downhill and no collision can occur there.
If particles move away from each other the potential difference increases if the particles is inside the attractive regime of their pair-potential.
Here a collision might occur.
The parts that contribute to an increase in the collision probability as computed from Eq.~\eqref{eq:P_no-coll} are illustrated in Fig.~\ref{fig:potentials_attractive}.

\begin{figure}
\includegraphics[width=1.0\columnwidth]{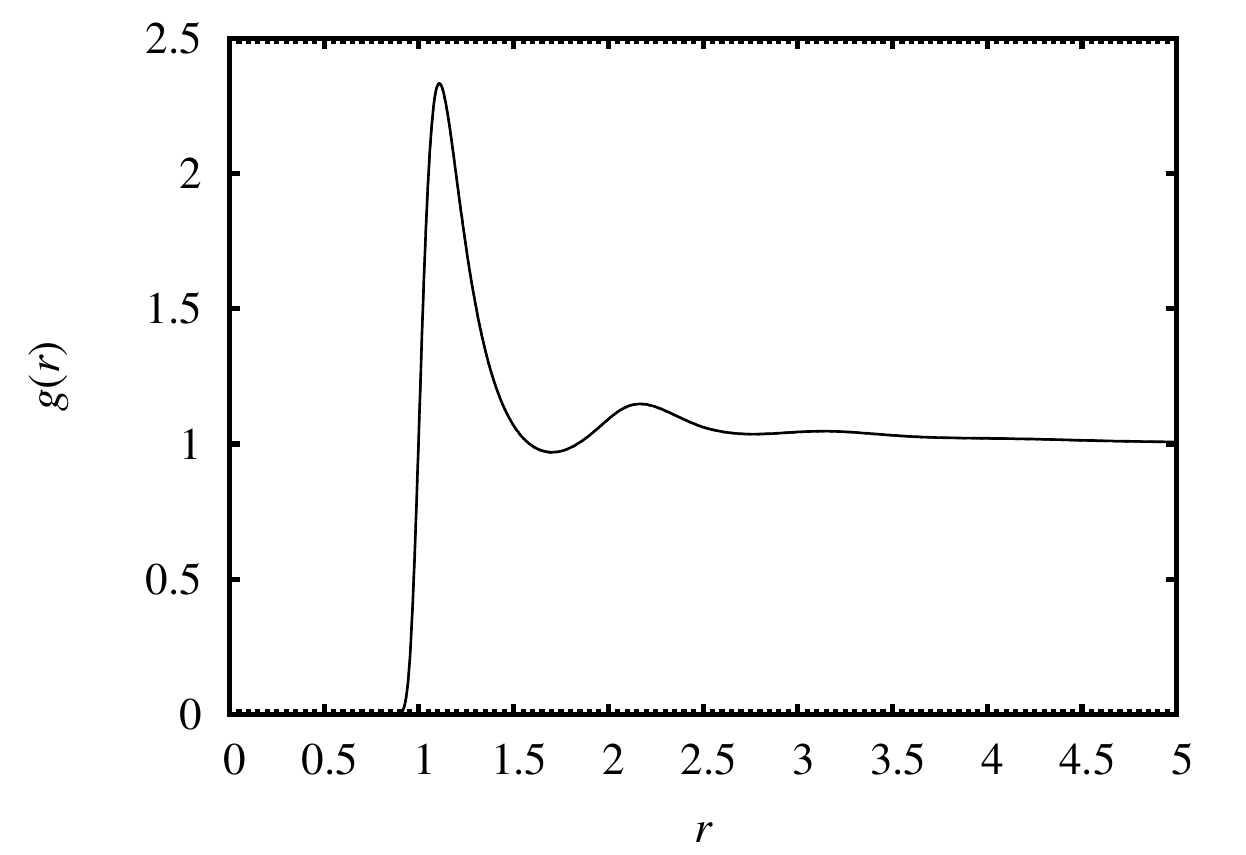}
\caption{The solid line shows the RDF of the truncated-shifted Lennard-Jones potential with $r_c =2.5$ at $\rho=0.317$ and $T=1.085$ (LJ-units). The curves computed with the rejection-free method and the Metropolis method are identical.}
\label{fig:RDF_LJ}
\end{figure}

Fig.~\ref{fig:RDF_LJ} shows the radial distribution function (RDF) for the density $\rho = 0.317$ and $T=1.085$ in LJ-units (and 1000 particles) for the case $r_c = 2.5 \sigma$.
This is the critical point of this truncated-shifted LJ-potential \cite{Smit92}.
We have chosen this point, instead of e.g. a liquid state point, because here there is a clear influence of the attractive part of the potential on the RDF.

This RDF has been determined with the event-driven scheme where all particles move simultaneously for two cases.
In the first case velocities are periodically redrawn from the correct Gaussian distribution.
In the second one the velocities are never reset.
We also implemented both a reversible and irreversible version of the straight event-chain method.
As a check the RDF was also computed using the Metropolis scheme.
All curves are identical within statistical errors.
The maximal absolute deviation amongst the presented curves is $0.006$ near $r=1.1$.

\section{Discussion}

The event-driven rejection-free MC method outlined in this paper was successfully applied to a Lennard-Jones fluid.
We only considered pair-potentials.
If one wants to simulate a molecular system also angle and torsion potentials need to be considered.
The collision rule such as defined by Eq.~\eqref{eq:collision_rule} can also be used for these kind of potentials.
In that case 3 or 4 particles are involved in a collision, but solving Eq.~\eqref{eq:collsion-time} will require some more computational effort.
The generalization of the straight event chain collisions to these kinds of potentials seems less trivial.

A priori it is not clear if for molecular systems the new method is less efficient than MD or not.
As demonstrated by the harmonic well example Fiq.~\ref{fig:harmonic}, the motion goes from one side of the potential to the other.
In MD one needs to resolve the oscillating motion by using sufficiently small time-steps.
The time that is won in this way can be spend on the more involved computation of computing events and maintaining the event-queue.
Although MD simulation tools are quite mature the algorithms for event-driven simulations are still being improved \cite{Miller04,Bannerman11}.
The method presented in this paper widens the realm of possible applications of the event-driven particle because a large class of potentials can be handled now.
It remains to be seen if the application of the method is suited for niche applications only, or that it can rival with MD and Metropolis-MC for general purpose molecular simulations.

\appendix
\section{Proof of correctness}
\label{app:proof}

In this appendix we will proof the validity of the rejection-free scheme.
We will do this by demonstrating that the canonical distribution is the invariant distribution of the dynamics of the system.

The probability distribution to have at time $s$ a particle system with positions $\mathbf{x}^n$ and velocities $\mathbf{v}^n$ is denoted by $\rho(\mathbf{x}^n, \mathbf{v}^n, s)$.
The total potential of the system is given by $\sum_\alpha U_\alpha$.
The probability to have not collided with a potential $U_\alpha$ is given by \eqref{eq:P_no-coll}.
The probability density per unit time to collide with potential $U_\alpha$ when in state $(\mathbf{x}^n, \mathbf{v}^n)$ equals
\begin{equation}\label{eq:collision_pdf}
\begin{split}
    p_\mathrm{coll} &= - \frac{d}{ds} P_{\mathrm{no-coll },\alpha}(s) = \beta \, \mathrm{max} \Bigl (\frac{d}{ds} U_\alpha(\mathbf{x}^n(s)), 0 \Bigr)\\
    &= \beta \, \mathrm{max} \Bigl (\mathbf{v}^n \cdot \boldsymbol{\nabla} U_\alpha, 0 \Bigr).
\end{split}
\end{equation}
Upon collision the velocity changes according to \eqref{eq:collision_rule}.
As a shorthand notation for the collision operator we will use $\mathbf{R}_\alpha=(\mathbf{I} - 2 \mathbf{P}_\alpha)$.
Two relevant properties of this operator are
\begin{equation}\label{eq:relection_operator}
    \mathbf{R}_\alpha \cdot \mathbf{R}_\alpha = \mathbf{I} \text{ and } \mathbf{R}_\alpha^\mathrm{T} \cdot \boldsymbol{\nabla} U_\alpha = - \boldsymbol{\nabla} U_\alpha.
\end{equation}
After a collision the velocities, $\mathbf{v}^n$, become $\mathbf{R}_\alpha \cdot \mathbf{v}^n$ and, vise-versa, $\mathbf{R}_\alpha \cdot \mathbf{v}^n$ changes into $\mathbf{v}^n$.

The change of $\rho(\mathbf{x}^n, \mathbf{v}^n, s)$ with time has three contributions: streaming, creation of states with velocities $\mathbf{v}^n$ due to collisions and annihilation of states with velocities $\mathbf{v}^n$,
\begin{multline}\label{eq:equation_of_change}
    \frac{\partial}{\partial s} \rho(\mathbf{x}^n, \mathbf{v}^n, s) = - \mathbf{v}^n \cdot \boldsymbol{\nabla} \rho \\
    + \beta \sum_\alpha \mathrm{max} \Bigl ( \mathbf{v}^n \cdot \mathbf{R}_\alpha^\mathrm{T} \cdot \boldsymbol{\nabla} U_\alpha, 0 \Bigr) \, \rho(\mathbf{x}^n, \mathbf{R}_\alpha \cdot \mathbf{v}^n, s)\\
 - \beta \sum_\alpha \mathrm{max} \Bigl (\mathbf{v}^n \cdot \boldsymbol{\nabla} U_\alpha, 0 \Bigr) \, \rho(\mathbf{x}^n, \mathbf{v}^n, s)
\end{multline}
In this equation the gradient operator denotes differentiation towards positions only and not towards velocities.
From the second relation in \eqref{eq:relection_operator} we find that $\mathbf{v}^n \cdot \mathbf{R}_\alpha^\mathrm{T} \cdot \boldsymbol{\nabla} U_\alpha = - \mathbf{v}^n \cdot \boldsymbol{\nabla} U_\alpha$.
Furthermore, from the definition of the projection operator, \eqref{eq:projection}, one can derive that the scalar $(\mathbf{v}^n) \cdot \mathbf{M}^{-1} \cdot \mathbf{v}^n$ is an invariant of the collision operator $\mathbf{R}_\alpha$ for any $\alpha$.
Therefore, if we assume the form $\rho(\mathbf{x}^n, \mathbf{R}_\alpha \mathbf{v}^n, s) = \rho_x(\mathbf{x}^n, s) \, f(\mathbf{v}^n \cdot \mathbf{M}^{-1} \cdot \mathbf{v}^n)$, we find that for the collision terms of \eqref{eq:equation_of_change}
\begin{equation}
\begin{split}
&\quad \beta \sum_\alpha \mathrm{max} \Bigl ( \mathbf{v}^n \cdot \mathbf{R}_\alpha^\mathrm{T} \cdot \boldsymbol{\nabla} U_\alpha, 0 \Bigr ) \, \rho(\mathbf{x}^n, \mathbf{R}_\alpha \cdot \mathbf{v}^n, s)\\
& \qquad - \beta \sum_\alpha \mathrm{max} \Bigl (\mathbf{v}^n \cdot \boldsymbol{\nabla} U_\alpha, 0 \Bigr) \, \rho(\mathbf{x}^n, \mathbf{v}^n, s)\\
&= \beta \sum_\alpha \Bigl( \mathrm{max}\Bigl(-\mathbf{v}^n \cdot \boldsymbol{\nabla} U_\alpha, 0 \Bigr) - \mathrm{max} \Bigl(\mathbf{v}^n \cdot \boldsymbol{\nabla} U_\alpha, 0 \Bigr) \Bigr) \rho_x \cdot f\\
&= - \beta \Bigl ( \sum_\alpha \mathbf{v}^n \cdot \boldsymbol{\nabla} U_\alpha \Bigr ) \, \rho_x \cdot f.
\end{split}
\end{equation}
Using this result we find that for a canonical distribution, $\rho_x = Z^{-1} \, \exp[-\beta \, \sum_\alpha U_\alpha]$, the streaming part and the collision terms in  \eqref{eq:equation_of_change} cancel.
This concludes the proof that the configurational canonical ensemble is indeed an invariant distribution of the dynamics generated by the rejection-free method.

%\bibliography{event-driven}


\end{document}